\def\citen#1{%
\edef\@tempa{\@ignspaftercomma,#1, \@end, }
\edef\@tempa{\expandafter\@ignendcommas\@tempa\@end}%
\if@filesw \immediate \write \@auxout {\string \citation {\@tempa}}\fi
\@tempcntb\m@ne \let\@h@ld\relax \def\@citea{}%
\@for \@citeb:=\@tempa\do {\@cmpresscites}%
\@h@ld}
\def\@ignspaftercomma#1, {\ifx\@end#1\@empty\else
   #1,\expandafter\@ignspaftercomma\fi}
\def\@ignendcommas,#1,\@end{#1}
\def\@cmpresscites{%
 \expandafter\let \expandafter\@B@citeB \csname b@\@citeb \endcsname
 \ifx\@B@citeB\relax 
    \@h@ld\@citea\@tempcntb\m@ne{\bf ?}%
    \@warning {Citation `\@citeb ' on page \thepage \space undefined}%
 \else
    \@tempcnta\@tempcntb \advance\@tempcnta\@ne
    \setbox\z@\hbox\bgroup 
    \ifnum0<0\@B@citeB \relax
       \egroup \@tempcntb\@B@citeB \relax
       \else \egroup \@tempcntb\m@ne \fi
    \ifnum\@tempcnta=\@tempcntb 
       \ifx\@h@ld\relax 
          \edef \@h@ld{\@citea\@B@citeB }%
       \else 
          \edef\@h@ld{\hbox{--}\penalty\@highpenalty
            \@B@citeB }%
       \fi
    \else   
       \@h@ld\@citea\@B@citeB
       \let\@h@ld\relax
 \fi\fi%
 \def\@citea{,\penalty\@highpenalty\hskip.13em plus.1em minus.1em}%
}
\def\cite{\leavevmode\unskip\@ifnextchar[{\@tempswatrue\@citew}%
            {\@tempswafalse\@citex}}
\def\@citew[#1]#2{\ifnum\lastpenalty=\z@ \penalty\@highpenalty \fi
   \ [{\multiply\@highpenalty 3 
   \citen{#2}},\penalty\@highpenalty\ #1]\spacefactor\@m}
\def\@citex#1{\begingroup \leavevmode \@tempcnta\@m \unskip
  \def\@tempa{\@cite{\citen{#1}}\spacefactor\@tempcnta\endgroup}%
  \futurelet\@tempb\@citey}%
\def\@citey{\let\@tempc\@tempa
   \ifx\@tempb.\ifnum\spacefactor>2999 \let\@tempb\relax\fi\let\@tempc\@citez
   \else\ifx\@tempb,\let\@tempc\@citez
   \else\ifx\@tempb:\let\@tempc\@citez 
   \else\ifx\@tempb;\let\@tempc\@citez 
   \fi\fi\fi\fi
   \@tempc}%
\def\@citez#1{\@tempcnta\sfcode`#1\@tempb\futurelet\@tempb\@citey}%
\def\@cite#1{$\m@th\the\scriptfont\z@\edef\bf{\the\scriptfont\bffam}%
      ^{\hbox{#1}}$}

\def\nopagenumbers{\pagestyle{empty}}              
    \nopagenumbers   

\def\pagesetup{
\textwidth 6.0in
\textheight 8.5in
\pagestyle{empty}
\topmargin -0.25truein
\oddsidemargin 0.30truein
\evensidemargin 0.30truein
\raggedbottom\parindent=20pt
\baselineskip=14pt}

\pagesetup

\def\@normalsize{\@setsize\normalsize{14pt}\xiipt\@xiipt
\abovedisplayskip 12pt plus3pt minus7pt%
\belowdisplayskip \abovedisplayskip
\abovedisplayshortskip  \z@ plus3pt%
\belowdisplayshortskip  6.5pt plus3.5pt minus3pt%
\let\@listi\@listI}   

\def\small{\@setsize\small{12pt}\xpt\@xpt
\abovedisplayskip 10pt plus2pt minus5pt%
\belowdisplayskip \abovedisplayskip
\abovedisplayshortskip  \z@ plus3pt%
\belowdisplayshortskip  6pt plus3pt minus3pt
\def\@listi{\leftmargin\leftmargini 
\topsep 6pt plus 2pt minus 2pt\parsep 3pt plus 2pt minus 1pt
\itemsep \parsep}}

\def\footnotesize{\@setsize\footnotesize{11pt}\ixpt\@ixpt
\abovedisplayskip 8.5pt plus 3pt minus 4pt%
\belowdisplayskip \abovedisplayskip
\abovedisplayshortskip \z@ plus2pt%
\belowdisplayshortskip 4pt plus2pt minus 2pt
\def\@listi{\leftmargin\leftmargini 
\topsep 4pt plus 2pt minus 2pt\parsep 2pt plus 1pt minus 1pt
\itemsep \parsep}}

\newcommand{\symbolfootnotes}{\renewcommand{\thefootnote}
	{\fnsymbol{footnote}}}

\newcommand{\alphafootnotes}
	{\setcounter{footnote}{0}
	 \renewcommand{\thefootnote}{\alph{footnote}}}

\newenvironment{Titlepage}{\parsep=0pt \topsep=0pt \symbolfootnotes}{\relax}

\def\Title#1\endTitle{\begin{center}%
   \baselineskip=16pt 
 \bf #1\\[.5cm]}                    
\def\Author#1#2\endAuthor{\small\it
   {\rm #1}\\[1pt] #2\\[.3cm]}
\def\endAuthors{\end{center}
                \vglue .5cm    
                      \alphafootnotes}

\newenvironment{Abstract}%
       {\centering\bgroup
          \begin{minipage}{30pc}\small
            \noindent
	    \centerline{\tenrm ABSTRACT}
	    \parindent=0pt}%
         {\end{minipage}\egroup\par
         \normalsize}

\def\section{\@startsection{section}{1}
{\z@}{.5truecm plus -0.5ex minus -0.1ex}{\medskipamount}{\bf}}
\def\subsection{\@startsection{subsection}{2}
{\z@}{\bigskipamount}{\smallskipamount}{\it}}
\def\subsubsection{\@startsection{subsubsection}{3}
{\z@}{\bigskipamount}{\smallskipamount}{\rm}}

\def\thesection       {\arabic{section}.}

\def\appendix{\par
  \setcounter{section}{0}
  \setcounter{subsection}{0}
  \def\thesection{\Alph{section}.}}

\def\@biblabel#1{#1.\hfill}

\def\thebibliography#1{\section*{References\@mkboth
  {\uppercase{References}}{\uppercase{References}}}\list
  {\@biblabel{\arabic{enumi}}}{\settowidth\labelwidth{\@biblabel{#1}}%
    \leftmargin\labelwidth
    \advance\leftmargin\labelsep
    \parsep\z@ plus\p@ \itemsep\z@ plus\p@ minus\p@ \topsep\z@
    \usecounter{enumi}%
    \let\p@enumi\@empty
    \def\theenumi{\arabic{enumi}}}%
    \def\newblock{\hskip .11em plus.33em minus.07em}%
    \sloppy\clubpenalty800\widowpenalty800
    \sfcode`\.=1000\relax
}

\long\def\@caption#1[#2]#3{\par\addcontentsline{\csname
  ext@#1\endcsname}{#1}{\protect\numberline{\csname
  the#1\endcsname}{\ignorespaces #2}}\begingroup
    \@parboxrestore
    \small
    \@makecaption{\csname fnum@#1\endcsname}{\ignorespaces #3}\par
  \endgroup}

\def\beq{\begin{equation}} \def\eeq{\end{equation}}  
\def\beqa{\begin{eqnarray}} \def\eeqa{\end{eqnarray}}

\chardef\bslash=`\\    

\documentstyle[12pt]{article}

\begin{document}

\begin{Titlepage}
\rightline{UCSBTH-94-38}
\rightline{CGPG-94/9-1}
\rightline{gr-qc/9409036}

\Title
The Spectral Analysis Inner Product for Quantum Gravity
\endTitle
\Author{DONALD MAROLF}
Department of Physics, The University of California,
Santa Barbara, CA 93106, USA
\endAuthor
\endAuthors

\begin{Abstract}
This submission to the Proceedings of the Seventh Marcel-Grossman
Conference is an advertisement for the use of the ``spectral
analysis inner product" for minisuperspace models in quantum
gravity.
\end{Abstract}
\end{Titlepage}

\def\o{\overline}
\section*{}

The following is intended to be an {\it advertisement}
for what will be called the
``spectral analysis construction" of a physical Hilbert space for cosmological
models.  It contains no proofs, but instead describes the
problem to be addressed and the spectral analysis
``solution" with references to the literature.

Our aim is to address Dirac quantization of homogeneous
cosmological models or, more generally, of time reparametrization
invariant systems with a finite
number of degrees of freedom.
When such a system is
presented in canonical form, it is described by a Hamiltonian
constraint $h=0$ on some phase space.  We use lower case
letters to represent classical functions on this phase space while
capital letters will represent quantum operators.

The Dirac quantization prescription states that the classical constraint
function $h$ should
be replaced by a quantum operator $H$ that acts on some linear
space ${\cal H}_{aux}$ of ``states," or ``wavefunctions."
The quantum version of the constraint is to be the selection of so-called
``physical states" $\psi_{phys}$ through the condition
$$H \psi_{phys} = 0.$$
Of course, ambiguities remain in this proposal since, for example, the
structure (Hilbert space, Banach space, etc.) of the ``auxiliary
space" ${\cal H}_{aux}$ on which $H$ acts has not been specified.
Also, it is not clear that it is strictly necessary that the
physical states be members of ${\cal H}_{aux}$ (this point will be
illustrated shortly).

In addition, the so called ``physical operators" should preserve the
space of physical states.  This is guaranteed if every
physical operator $A$ is
required to commute with the
constraint operator $H$.  This is the quantum version of the
statement that a classical gauge invariant ($a$) has vanishing Poisson
bracket with the constraint, $\{a,h\} = 0$.

By analogy with quantum mechanics of unconstrained systems, we would
like our physical states to form a Hilbert space.  The inner product
on this space is not specified by the Dirac prescription and its
specification is exactly the problem that we wish to address here.
Since this question has been discussed a number of times and various proposals
have been made, we quickly mention a few popular ones
and comment on their properties before turning to the spectral analysis
approach itself.

One such proposal is known as the Klein-Gordon inner product and
is motivated by the fact that, in a common presentation, the constraint
$H \psi_{phys} = 0$ takes the form of a Klein-Gordon equation with a potential.
Difficulties arise, however, because the potential is usually
``time-dependent."  While Wald$^1$ has shown that a form of
Klein-Gordon inner product can still be defined in such cases and that
it is conserved, the corresponding quantum cosmologies expand forever$^1$
even when the corresponding classical models reach a maximum
volume and then recollapse.  Such models also display
additional unusual behavior at this turning point$^1$.

A second  proposal, called the ``deparametrization" proposal, asks that
a canonical transformation be performed
that renders the constraint linear in some momentum $P_0$.  This
form is associated with deparametrization of the model using
the conjugate variable $x^0$ as an internal clock.
If the quantum $H$ is constructed by replacing $P_0$ with $-i{ {\partial}
\over {\partial x^0}}$, then the quantum constraint takes the form of a
Schr\"odinger equation.  As in
unconstrained quantum mechanics, some sort of $L^2$ inner product can
usually then be imposed.  Unfortunately,
such a canonical transformation is not always apparent and has not been
shown to exist for a generic model.

In addition to noting these technical difficulties,
we object in-principle to the
choice of a certain
classical phase space function for special treatment as a ``time" variable
in these approaches.
The philosophy followed in {\it this} work and in the references is that
all degrees of freedom are to be treated on an equal footing
\footnote{On the other hand, we have no such objection to, for example, the
algebraic program of Ref. 3 and the spectral analysis approach can be
used to supplement such ideas.}.

The spectral analysis proposal gives a construction of an inner
product for the physical Hilbert space of finite dimensional
time reparametrization invariant quantum theories for which
the quantum constraint is presented in the form $H \psi = 0$
where $H$ is an operator in some Hilbert space and zero lies
in it's continuous spectrum.  This inner product was proposed
independently in Refs. 4,5,6 and Ref. 7 (in the context of
linearized quantum gravity) and
was in fact used in Ref. 7 in a more general form
appropriate to the presence of an arbitrary number of constraints that
form a Lie algebra.
We will use a presentation along the
lines of Refs. 4 and 5 as it is in these works that the
properties of the approach claimed below are derived.

We first introduce the spectral analysis inner product by analogy.
Suppose that the spectrum of our constraint operator ($H$)
is entirely discrete.  Then, there is a natural inner product
to use on the solutions of $H\psi = 0$; namely,
just the inner product induced from the auxiliary space
${\cal H}$.  However, let us introduce this inner product on
${\cal H}_{phys}$ in a slightly different way.  We note that
the spectral theorem allows us to write the auxiliary
Hilbert space uniquely in the form ${\cal H}_{aux} = \oplus_{\lambda
\in \sigma(H)} {\cal H}_{\lambda}$ where each ${\cal H}_{\lambda}$
contains the eigenvectors of $H$ with eigenvalue $\lambda$.
The physical Hilbert space may thus be defined to be
${\cal H}_0$.  Operators that commute with $H$ will have a natural
induced action on this Hilbert space which is Hermitian if the
operator is Hermitian on ${\cal H}_{aux}$.

The spectral analysis proposal is based on the observation that a
similar fact is true in the case where $H$ has purely continuous
spectrum.  That is, at least when the spectrum of $H$ is
``uniform" (see Ref. 5) near zero, the auxiliary space
can be written as
\begin{equation}
{\cal H}_{aux} = \oplus \int_{\lambda \in \sigma(H)} d\lambda \
{\cal H}_{\lambda}
\end{equation}
where, in fact, ${\cal H}_{\lambda}$ and ${\cal H}_{\lambda'}$
are isomorphic
for $\lambda, \lambda'$ sufficiently close to zero.  Formally, at
least, this decomposition is unique so that we may define
${\cal H}_{phys} \equiv {\cal H}_0$.  Now, ``sufficiently
smooth" operators have a natural induced action on ${\cal H}_{phys}$
and this action is symmetric if the original operators were
symmetric in ${\cal H}_{aux}$.

The following is a list of claims for the spectral analysis
approach that are verified in Refs. 4 and 5.

1)  This method defines a positive definite inner product.

2)  Spectral analysis treats all degrees of freedom on an equal footing.

3)  The method can be applied to all (Hamiltonian) Bianchi models,
including Bianchi IX.

4)  An overcomplete set of quantum observables (perennials in the
terminology of Ref. 11) can be constructed for these models
that act as hermitian operators in the Hilbert space defined
by the spectral analysis inner product.

5)  The spectral analysis results coincide with the usual results in the
case of parameterized Newtonian systems.

6)  In a spectral analysis quantization of recollapsing cosmological
models, the quantum cosmologies also recollapse.

This concludes our advertisement for the spectral analysis approach.  It
is hoped the the results of Refs. 4 and 5 will be extended
in the near future.

\section*{Acknowledgements}

This work was partially supported by NSF grants PHY93-96246,
PHY90-08502 and by funds from the Pennsylvania State University.

\end{document}